# 3D Printing of Flexible Room-Temperature Liquid Metal Battery

**Fujun Liu[1], Yongze Yu[1], Lei Wang[1], Liting Yi[1], Jinrong Lu[1], Bin Yuan[1,2], Sicong Tan[1,2], and Jing Liu,[1,2,3]***

[1]Technical Institute of Physics and Chemistry, Chinese Academy of Sciences, Beijing 100190, China.

[2]School of Future Technology, University of Chinese Academy of Sciences, Beijing 100049, China.

[3]Department of Biomedical Engineering, School of Medicine, Tsinghua University, Beijing 100084, China.

*Corresponding author. Email: jliu@mail.ipc.ac.cn.

**Abstract:** The entirely soft and transformable room-temperature liquid metal battery based on 3D hybrid printing was proposed and experimentally demonstrated. Liquid metal gallium and conductive gel were employed as the negative and positive electrode material, respectively. The discharge process of batteries with different structure, single-cell and multi-cell, were clarified. The results indicated that the obtained battery possessed rather stable and persisted discharge property, but the multi-cell battery presented stair-step decrease overtime. SEM and EDS images were obtained to evaluate the morphological change of the electrode during the discharge process, and the liquid metal was found to be easily exposed and active according to the solubility of the alkaline electrolyte to oxidized gallium. The successful 3D hybrid printing allowed different polymers, including PLA, ABS and TPU, to be precisely patterned in any required battery body structures, with liquid electrodes in appropriate specified positions. The flexibility and practicability of this new generation battery was illustrated through working devices. The present soft battery offers important power supply for future wearable and epidermal electronics.

**Keywords:** Liquid metal battery; 3D printing; Soft machine; Flexible energy; Wearable electronics.

## 1. Introduction

Large-scale energy storage has been a worldwide hotspot for scientific research and industrial circles, including intermittent renewable resources [1]. Batteries are employed widely for energy storage due to their low cost, high energy density, simplicity and reliability[2]. However, as the continuous improvement of renewable energy technologies, including solar and wind power, has pressed for long-life and large-scale energy storage, conventional battery technologies cannot meet with rising requirements[3]. Through comparison with the rapid development of high-performance Li-ion batteries for portable electronics[4], liquid-layer galvanic cell, also called as all-liquid energy storage cell or liquid metal battery[5], better suits new high-energy-density demands, like soft electronics, automotive applications etc. The history of liquid battery, in which a great quantity of electricity could be stored, initiated from the aluminum industry, extracting aluminum from aluminum-copper alloy, known as the Hoops procedure[6]. Being instead the story of a society catching up with a technology far ahead of its age, liquid batteries enable large current density (larger than 0.5 A/cm$^2$) and large energy storage (~3 kAh of electrical charges from the oxidation of 1 kg of aluminum)[7]. Sadoway and his cooperators registered their first patent of liquid metal battery in 2008,[8] followed by publications of different combinations of metals and electrolytes in a form of liquid metal batteries[9], which focused primarily on active metals of 1$^{st}$ and 2$^{nd}$ groups in the periodic table. In their outstanding technologies, very promising energy storage capability was presented. An only pity is that such technology is basically a rigid system due to application of high melting point metal and their running is somewhat difficult. This may restrict such battery's practical value. From an alternative, in the year of 2015, the present lab proposed and demonstrated an entirely soft liquid metal battery through introduction of the matching room temperature liquid metals and alloy, as well as the allied electrolyte [5]. Over the continuous endeavor, a group of different fabrication and combination strategies regarding the testing of the electronic materials and the packaging soft materials including applications were investigated and clarified. Part of the progresses has also been presented in international conferences.

Overall, a liquid metal battery basically consists of two liquid metal electrodes, separated by electrolyte of molten salt, which will spontaneously segregate into three layers based on the immiscibility of each component (as shown in Fig .1). Generally there are two types of such battery, depending on the transport carrier of electric current in the electrolyte. One is based on two uncorrelated half-cell reactions with charge transfer on electrons, as shown in Fig. 1-a and Fig. 1-b, describing the discharging and charging procedures, respectively. And the other type is based on the binary alloy electrode combinations, as shown in Fig. 1-c and Fig. 1-d where ionized B from anode transfers through the electrolyte to form alloy with the cathode material in the discharging process. All current reports on liquid metal generally focus on binary-alloy-electrode batteries, the main advantages of which are super kinetics and transport properties, including extremely high ionic conductivity of the electrolyte, resistance of individual liquid phases (with no structural defects, dendrites, cavities, and so on) and good contact between the liquid phases[10]. Meanwhile, liquid metal batteries are also low-cost, as most of the candidate electrode materials (including aluminum, magnesium, stibonium, and so on) are earth-abundant and inexpensive. However, liquid metal batteries possess some disadvantages, the primary one of which relates to the high working temperature (generally >200 $^o$C, 930-950 $^o$C for aluminum electrolysis, 700 °C for magnesium electrolysis, etc.), making them unsuitable for practical applications. Also the fast corrosion limited this kind of liquid metal battery to stationary applications only. Therefore, room-temperature liquid metal batteries are urgently needed for the energy storage [5].

Gallium is one of the room-temperature liquid metals, recently attracting widespread attentions due to its potential applications in heat transfer[11] and electronic circuit printing[12]. Due to its negligible vapor pressure, low toxicity, low melting point (29 °C in standard atmosphere) and appropriate oxidation-reduction potential ($E^0 Ga^{3+}/Ga=-0.560\pm0.005v$)[13], gallium can be used as the electrode material in novel liquid metal battery. When exposed to air, gallium often forms a thin (several nanometers) passivating "skin" on the surface, composed of gallium oxide[14], which can be easily dissolved by NaOH or KOH solution. Therefore, gallium could be used as the electrode material and KOH solution as the electrolyte.

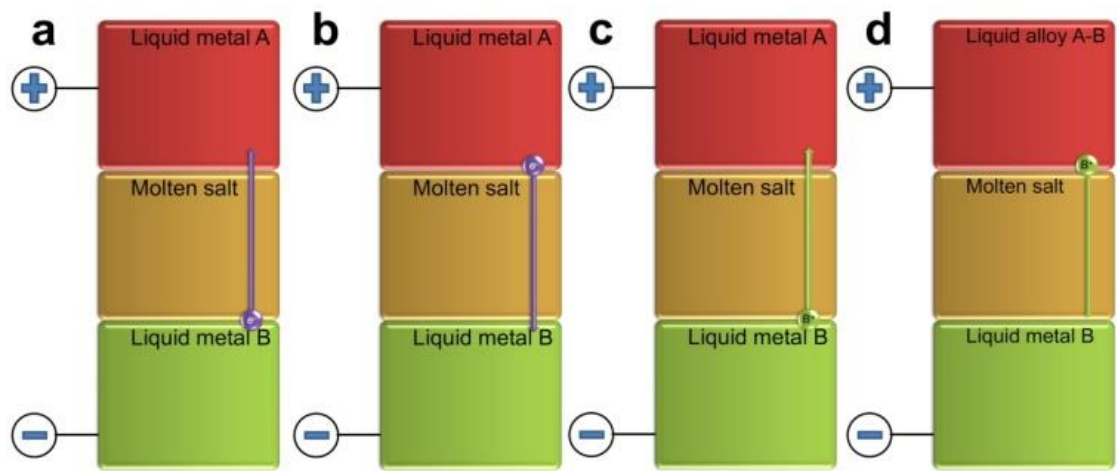

**Fig. 1.** Schematic diagram of an electro-transfer type liquid metal battery upon a) discharging and b) charging, and ion-transfer type one upon c) discharging and d) charging.

3D printing, or additive manufacturing (AM), is a now technology that is expected to up-end traditional approaches to design and manufacture, with profound economic, social, geopolitical, environmental and security meanings[15]. In brief, product by 3D printing can be "printed" or built layer-by-layer and additively, rather than by subtracting larger block of material into smaller elements, which is subtractive manufacturing. Traditional notion of printing involves transferring ink into characters or patterns on paper, line-by-line until the documents is completed. Additionally, another parameter, the height *h*, was led into 3D printing, where "ink" (with enormous materials included, even cells for printing of tissues[16]) is printed layer-by-layer until the product is completed. The principle of this newly emerging technology is using computer data sets of 3D models to reconstruct a 3D physical model by the addition of successive layers of liquid, powder, or the sheet material, which can create objects with almost any complex shape or geometric feature, as the primary advantage of this technology.

With the rapid development of wearable electronics and epidermal electronics, urgent request was on the fabrication of flexible batteries with outstanding electronic properties, including stable discharge, considerable capacity, good safety, and even favorable transparency for some devices[17]. Most of current reports on flexible battery were focused on using elastic polymer as the cell body, or doping powder-form electrode materials into flexible polymer, which may degrade the cell performance and make it unsuited for some applications. Sadoway's liquid metal batteries can avoid these defects as liquidity endows them with superior kinetics and transports properties, along with low cost according to the earth-abundant and inexpensive candidate electrode materials. However, the elevated operating temperature (usually higher than 200 $^{o}$C) due to use of high melting point metals limits the applications of such batteries. Therefore, in this article we turn to another completely different battery material system, say room temperature liquid gallium and conductive gel as the cathode and anode material, respectively, and KOH solution as the electrolyte. This leads to the realization of an entirely flexible room-temperature liquid metal battery, with elastic polymer as the cell body fabricated by 3D printing. Such soft energy system possesses big potential in personal wearable/epidermal electronics.

## 2. Materials and Methods

### 2.1 Fabrication of Flexible Liquid Metal Battery

Gallium (99.999%, purchased from Anhui Rare New Materials Co. LTD, Anhui, China) and electrode gel (SPECTRA 360, Parker Laboratories Inc., Fairfield, USA) were employed as the cathode and anode material, respectively, and thermoplastic urethane (TPU-1.75mm, Huiwei Printing Supplies Materials Inc., Shenzhen, China) was used as the cell body fabricated by 3D printer (self-assembled in our lab). The work temperature for the heating jet was 195 $^{o}$C, and relative softwares were Repetier-Host and Auto-CAD 2014. 0.3M NaOH solution was added into the cavity in the cell, which was then covered by transparent adhesive tape. The amounts of gallium and electrode gel used in a single battery were 0.12 and 0.04g, respectively, and used NaOH solution was 0.2mL.

## 2.2 Electrochemical Characterization of Prepared battery

The electrochemical properties of the tested battery were measured by setting up a simple electric circuit, in which two positive and negative charges of as-obtained liquid metal battery were connected to each end of a standard resistance of 200 Ω and a data acquisition (Agilent 34970A, Agilent Technologies, USA) was employed for real-time measurements of the voltage on the working resistance, until the voltage decreased to 1 % of the initial value.

## 2.3 Morphology Analysis of Liquid Metal Electrode

To study the changes on morphology and chemical constitution of the liquid metal electrode, before and after the discharge procedure, a Nova NanoSEM scanning electron microscope (SEM) with energy dispersive spectrometer (EDS) analysis (NOVA Nano SEM 430 +EDS, FEI Co., Hillaboro, USA).

## 2.4 Wettability Study of Liquid Metal on Polymers

The wettability study of liquid metal was conducted using a JC2000D3 goniometer (Powereach CO., Ltd, Shanghai, People's Republic of China) by sessile drop method. The contact angles measured by advancing or receding a small volume of liquid (~ 2 μL) onto the surface of polymer plate fabricated by 3D printer. At least three measurements were performed on each surface.

## 3. Results and Discussion

## 3.1 Electrochemical and Morphology Characterization

In liquid metal batteries, gallium acted as active metal that was oxidized and released electrons which passed from the negative electrode (gallium), via the load circuit, then to the positive electrode (silver wire) to form the electric current, when battery was discharging. The potential difference was created according to different chemical potentials of the liquid metal (Ga) in the negative electrode and water in the electrolyte, presented with the following cell diagram:

$$Ga_{(l)} \mid Ga^{3+}_{(dissolved\ in\ electrolyte)} \mid electrolyte_{(l)} \mid H_2O_{(l)} \mid H_{2(g)} \qquad (1)$$

Individual reactions at the electrodes can be written as:

$$\text{Negative electrode: } Ga = Ga^{3+} + 3e^- \qquad (2)$$

Positive electrode: $2H_2O + 2e^- = H_2 + 2OH^-$ (3)

Therefore, the total electrochemical reaction can be simply written as :

$$2Ga + 6H_2O = 2Ga^{3+} + 3H_2 + 6OH^-$$ (4)

where, Ga loses electrons to form $Ga^{3+}$ ion, and $H_2O$ gains electrons to form $H_2$ and $OH^-$ ions. The formed $H_2$ will gather in the interface between electrode gel and electrolyte, and remove out of the battery.

To study the electrochemical behavior of liquid metal of gallium, two kinds of batteries were investigated, one with 24 cells connected in series and a single one, as shown in Fig. 2-b and Fig. 2-c, respectively. The schematic of multi-cell batteries was shown in Fig. 2-a, where single cells were in series connection. By closely contact conductive gel to Ga electrode of another cell, the achieved load voltage can be as high as required. In discharge curves of the Ga-H system (Fig. 2-d), the initiative load voltages of single-cell and 24-cell batteries were 0.5 and 9.8 V, respectively, with open-circuit voltages of 1.1 and 25.8 V, respectively. Therefore, the initiative internal resistance of a single Ga-cell was predicted to be about 110-120 V, which was much higher than the average value of batteries on the market[18] but equivalent of some novel ones in research[19]. This significant internal resistant was attributed to the large electrode gap, which resulted in poor operating current capability and significant voltage drop under heavy load and will be overcome in the future work. By comparison, a distinct was obtained on shapes of the two discharge curves in Fig. 2-d. The voltage of the single cell did not decrease (the red curve in Fig. 2-i) over time during several days, which can be explained by the long-term stability of the gallium electrode. The SEM images in Fig. 3-a and Fig. 3-c presented the surface morphology of gallium before and after discharge process, respectively. It can be clearly observed that the discharge process did not cause any change on the morphology of negative electrode, with magnified details in Fig. 3-b and Fig. 3-d, which some wrinkles emerged according to the surface oxidation of gallium during the sample preparation process[20]. This oxidation led to a sudden and substantial drop in the surface tension[21], and the formed oxidation layer in solid state will prevent the electrochemical reaction of gallium due to the arrested contact with electrolyte, which happened in previous work when NaCl solution was used as the electrolyte and it resulted in the "death" of batteries. Therefore, the electrolyte in this article was replaced by NaOH solution, as the product of oxidation is soluble in alkaline solution[22]. In this region of pH, the reaction product should be $Ga_2O_3$ according to thermodynamic prediction, which forms Ga(OH)$_3$ hydrated complexes solvable in NaOH solution[23]. Due to the "cleaning power" of electrolyte, the surface of gallium electrode kept chemically active during the discharge process, which was further affirmed by the EDS spectra in Fig. 3-e and Fig. 3-f. The conductivity of the electrolyte NaOH solution was detected by electrochemical impedance spectroscopy (EIS), and the result was shown in Fig. 3-g. The forms of resistance- and phase- curves were similar with those of a simplified Randles cell, as real value of resistance is the sum of the polarization resistance and the solution resistance. The polarization resistance got decreased as the frequency is increasing.

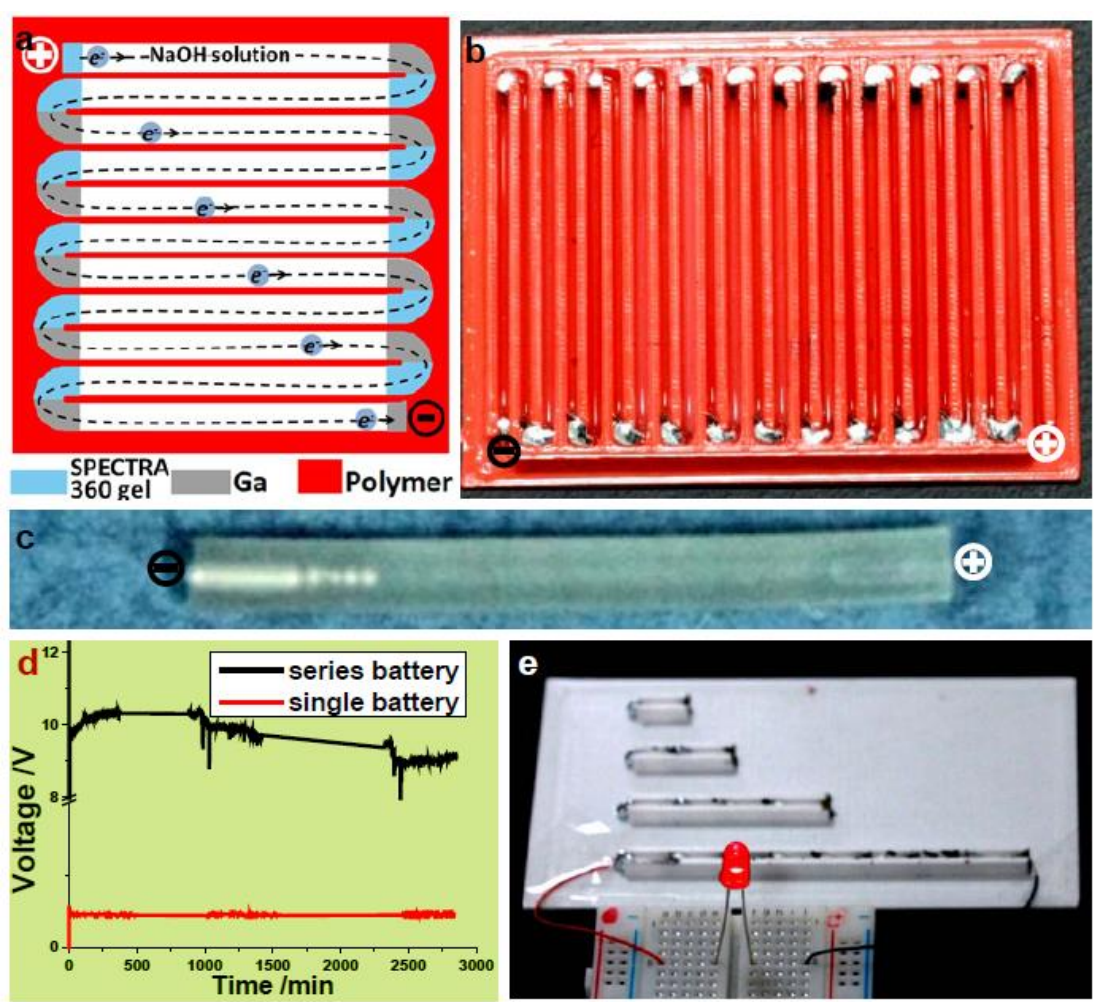

**Fig. 2.** Electrochemical study of liquid metal batteries. a) presents the schematic for battery with multiple cells, where the dotted line indicates the migration of electrons inside batteries during charging process; b) and c) present images of liquid metal batteries of 24-cell series and single one, respectively; d) presents discharge curves of the two batteries in b) and c), with load resistance of 98 and 2k ohm, respectively; e) a light-emitting diode (LED) can light up according to the electrical support from the liquid metal battery.

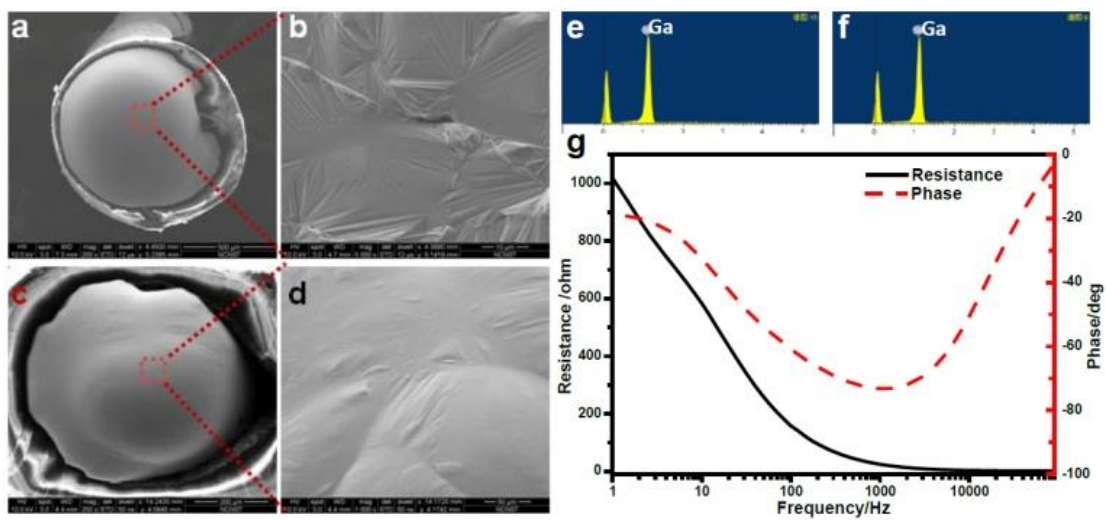

**Fig. 3.** Morphology analysis of electrode. a) and c) present the SEM images of the Ga electrode before and after discharging process with local zoom in detail shown in b) and d), respectively; e) and f) present EDS map showing 100% gallium concentrations in sectioned platelets; g) presents the electrochemical impedance spectroscopy (EIS) of NaOH solution as the electrolyte.

By comparison, there was no difference between the two EDS spectra before and after discharge process, in both of which only peak for the element gallium emerged (content being 100%). Meanwhile the voltage of 24-cell battery series presented stair-step decrease overtime (black curve in Fig. 2-d), and the reason for this phenomenon was conjectured to be the heterogeneity of the prepared battery pack and the inactivation of positive electrode due to hydrogen bubbles escaping from the electrolyte. When one of the cells goes wrong during discharge process, like movement of liquid electrolytes, the total voltage of battery pack will descended to a lower level[24]. The electrochemical reaction on the positive electrode can be described as Eq. 3, where hydronium ions were reduced to hydrogen $H_2$ bubbles[25]. In single-cell battery (Fig. 2-h), generated $H_2$ bubbles can escape from the open end of the battery, but in the series battery (Fig. 2-b), $H_2$ bubbles can not escape but only absorb on the surface of electrode, leading to the inactivation of the electrode and degradation of the battery. These defects are inevitable at the present stage of research, and should be studied and overcome in the future study. To intuitively show the capacity, a LED was connected to the liquid metal battery and lighted up for more than two weeks until obvious darkening was perceived (sown in Fig. 2-e).

To make further studies on electrochemical properties of fabricated battery, discharge curves under different currents were achieved and shown in Fig. 4. When small discharge currents were adopted (10 and 20mA), the voltage of battery showed a petty decline in 1 hour, and the alight LED can last for more than 4 hours. When the discharge current increased to 40mA, the voltage presented a fast and obvious decrease, from 0.4V to 0.15V in 1 hour. This indicates that the fabricated battery is very fit for low-current applications.

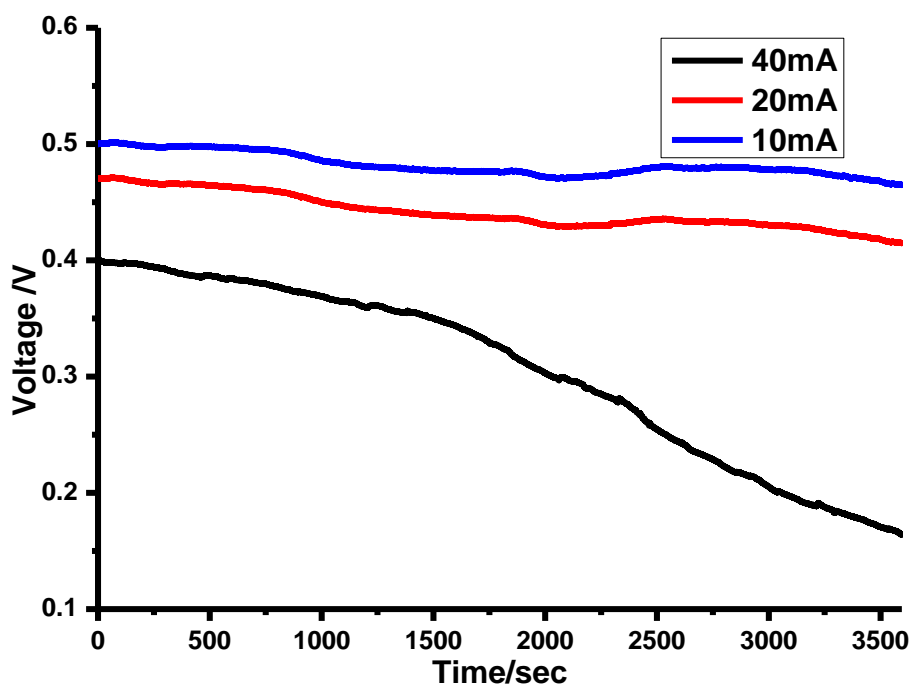

**Fig. 4.** Discharge curves of liquid metal battery under different discharge currents.

**3.2 Hybrid Printing of Batteries**

The technology of 3D printing had been used for fabrications of structural electronics due to allowing functional inks to be precisely patterned in filamentary architectures[26]. Our facile 3D printing technique allows functional inks, including liquid metal, polylactic acid (PLA), acrylonitrile-butadiene-styrene (ABS), thermoplastic polyurethane (TPU) and conductive gel, to be precisely patterned on many kinds of materials with flat surfaces. We harness this capability to fabricated 3D interdigitated battery architectures composed of liquid metal gallium, which serves as the negative electrode materials. The success of printing depends mostly on the integrity of the first layer printed, as the geometry of the first layer is critically important for the adhesion of ink to the substrate and subsequent deposition of ink on preformed object[27]. Therefore, a gracile double-sided tape was bonded on the aluminous platform to form a sticky substrate surface, on which melted ink can deposit to form thetic structure, shown in Fig. 5-a. Due to the simplification and flexibility of the technology of 3D printing, batteries of different structures were fabricated (see Fig. 2-g, Fig. 5-b and Fig. 5-c).

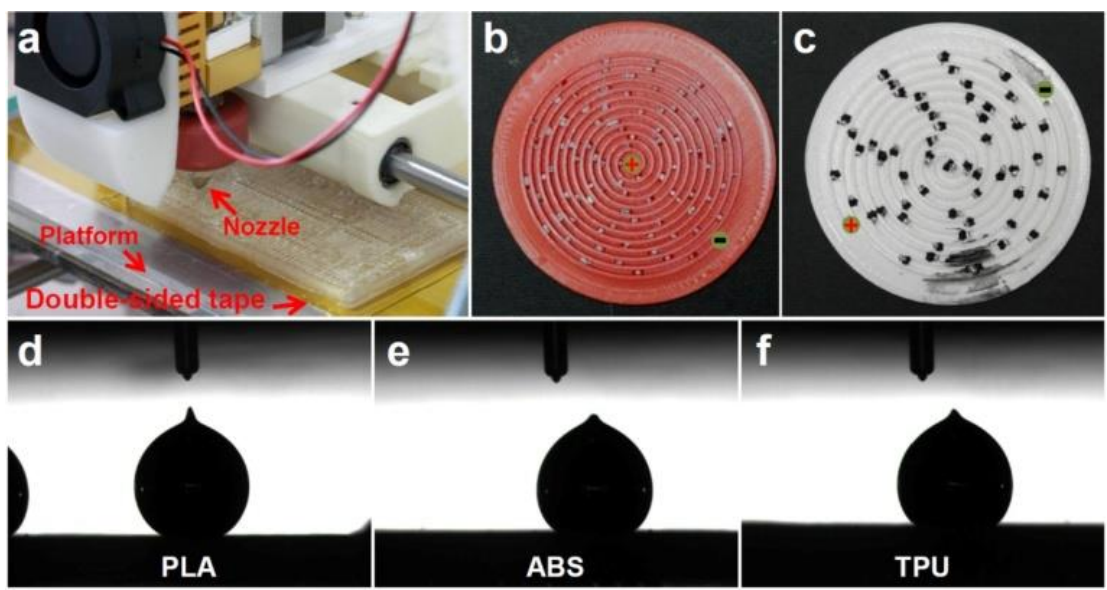

**Fig. 5.** Fabrication of liquid metal batteries. a) presents the operating state of hybrid printing; b) and c) give the structures of 89-cell and 54-cell batteries as representative of made-up batteries with different shapes, open circuit voltage being 98.2 V and 61.7 V, respectively; d), e) and f) show the wettability study of liquid metal on different polymer materials, PLA, ABS and TPU, for battery body building, with contact angle of $153^{o}$, $141^{o}$ and $140^{o}$, respectively.

As there were three kinds of inks in fabricating batteries, the most important factor for the hybrid printing should be the wettability between each component. The conductive gel, used as the positive electrode, can adhere to polymer effectively and remain its shape over a long period of time (transparent in Fig. 5-b and black in Fig. 5-c), so only the wettability between liquid metal and the polymer was studied here. The results were shown in Fig. 5-d, Fig. 5-e and Fig. 5-f, presenting the contact angles of liquid metal on PLA, ABS and TPU being $153^{o}$, $141^{o}$ and $140^{o}$, respectively, which indicated that liquid metal was non-wetting on polymer as the battery body. The non-wetting of liquid metal is very necessary and important for electrical properties of the battery, because the liquid metal electrode is required to keep in liquid state during discharge. If wetting on polymer, liquid metal will lay open on the bottom of cell body due to its heavy density, which will lead to the inactivation of batteries.

According to the non-wetness, liquid metal can remain a liquid drop in the cell body during discharge process, which plays an important role for the discharge stability of batteries.

**3.3 Flexible Epidermal Electronics**

The mostly potential application of this flexible battery is expected to power supply for wearable electronics, so it is very important for the battery to possess properties of tenacity, favorable biocompatibility and adjustable flexural strength. The polymer used in Fig. 2-b was hard and nontransparent, including PLA and ABS which can be used when superior mechanical strength and colorful product appearance were required. However, for some wearable electronics and epidermal electronics, batteries need to be bonded on the skin surface, which means the batteries would always suffer from tolerance, compression, stretch and even expansion[28]. Therefore, TPU was employed here to produce the epidermal batteries due to its biocompatibility and flexibility[29]. Fig. 6-a shows the structure of liquid metal battery made of TPU on 3M double-sided tape, which was packaged by cellulose tape when these cells were fulfilled with electrolyte. To display the structure more clearly, rhodamine solution was injected in the cells to obtain an fluorescent image, where 23 cells were connected end to end and positive electrodes (conductive gel) and negative electrodes (liquid metal) were dispersed in the corners of printed groove to achieve high aspect ratio architecture. The open-circuit voltage of this battery was measured to be 26.1 V, which was high enough for most of applied electronics such as those for electrocardiogram (ECG), electromyography (EMG) and blood pressure (BP) detection.

The bendability of obtained liquid metal battery was shown in Fig. 66-c, where the battery can be twisted in any shape under considerable force, due to the small thickness (2 mm) and great flexibility of TPU and electrodes, and no structural failure was observed after repeatedly bending and torture. After removed the red substratum, the battery can be bonder on the skin (Fig. 6-d). The tiny volume and small weight of the battery imposed negligible mechanical or mass loading (typical total mass of ~10 g), as is evident from the image of Fig. 6-d. On the surface of the skin, there are wrinkles, creases and pits with amplitudes and feature sizes of 15-100 μm and 40-1000 μm[30], respectively. The liquid metal battery produced here presented thickness and other physical properties that were well matched to the epidermis, with the ability to conform to the relief on its surface. Therefore, we refer to this battery as outstanding power supply for epidermal electronics.

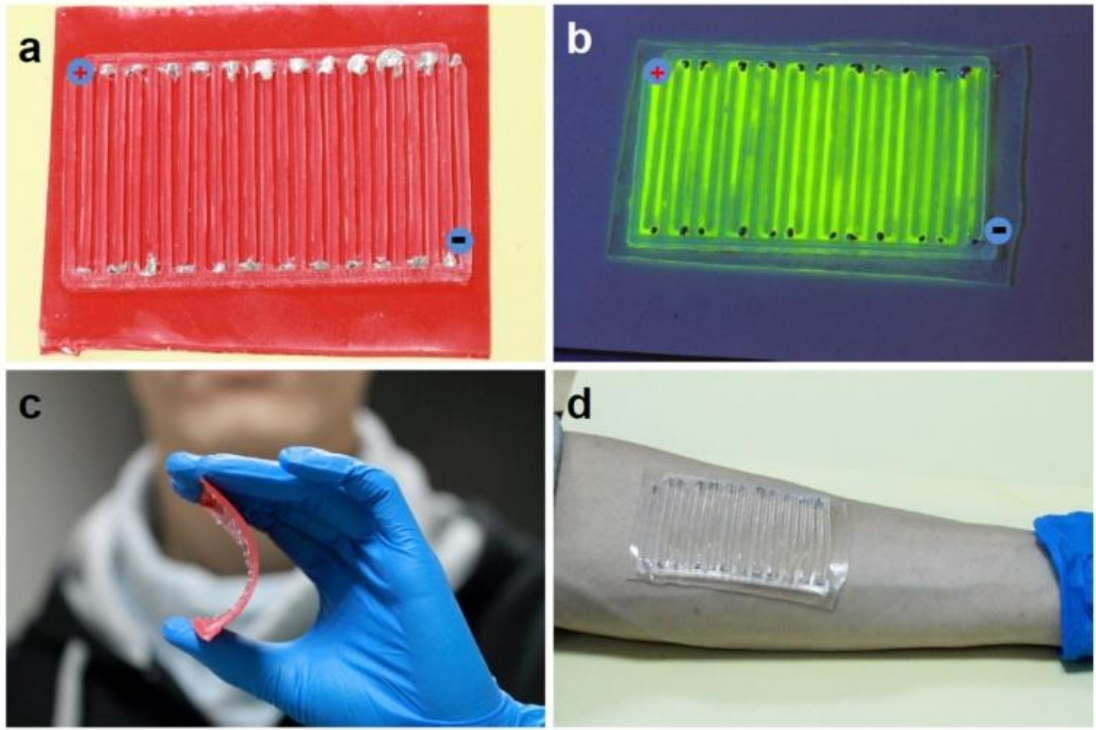

**Fig. 6.** a) Flexible battery was fabricated by TPU on 3M soft double-sided tape; b) Rhodamine solution was used to fulfill the cells and the picture was taken under UV lamp to clearly show the structure of cell body; c) Fabricated battery was flexible and can be twisted in any shape; d) Flexible liquid metal battery on skin: after removed the red substratum, the flexible and transparent battery can be bonded anywhere on the skin, according to applications.

4. **Conclusion**

We have successfully demonstrated that it is possible to prepare various types of operating liquid metal batteries on the basis of 3D hybrid printing. The non-wetting of liquid metal on polymer cell body is quite necessary and important for stable discharge process. Although the discharge of the single cell had been successfully confirmed to be stable and persistent according to the "cleaning power" of alkaline electrolyte which kept the liquid metal electrode active, it needs to be pointed out that the load voltage of multi-cell battery series presented stair-step decrease overtime, due to the heterogeneity of battery pack and the inactivation of positive electrode by hydrogen bubbles. This disadvantage would be overcome in the future research. When TPU was employed as the battery body material, the produced battery was confirmed to be appropriate for power supply to epidermal electronics, as obtained battery can be bonded anywhere on the skin surface.

**Acknowledgment**

This work is partially supported by the NSFC under Grant 91748206, Dean's Research Funding and the Key Research Program of Frontier Sciences of the Chinese Academy of Sciences.**References**

1. J. Wang and S. Kaskel, S. J. Mater. Chem., 2012, 22, 23710-23725; C. Lai, M. Lu and L. Chen, J. Mater. Chem., 2012, 22, 19-30.
2. M. Duduta, B. Ho, V. C. Wood V C, et al. Adv. Energy Mater., 2011, 1, 511-516; F. M. Hassan, R. Batmaz, J. Li, et al. Nat. Commun. 2015, 6, 8597; F. M. Hassan, V. Chabot, A. R. Elsayed, et al. Nano Lett., 2013, 14, 277-283.
3. C. Budischak, D. A. Sewell, H. Thomson, et al. J. Power Sources, 2013, 225, 60-74; F. Díaz-González, A. Sumper, O. Gomis-Bellmunt, et al. Renew. Sust. Energy Rev., 2012, 16, 2154-2171.
4. Z. Wang, T. Chen, W. Chen, et al. J. Mater. Chem. A, 2013, 1, 2202-2210; Y. Shi, S. Chou, J. Wang, et al. J. Mater. Chem. A, 2012, 22, 16465-16470.
5. L. Wang, J. Liu, Liquid Metal Soft Battery and Its Fabrication Method, China Patent: CN201510689223.8, 2015-10-21; L. Wang, J. Liu, A Room Temperature Liquid Metal Battery, China Patent: CN201610573221.7, 2016-7-20; D. J. Bradwell, H. Kim, A. H. C. Sirk, et al. J. Amer. Chem. Soc., 2012, 134, 1895-1897; G. H. Miley, N. Luo N, Mather J, et al. J. Power Sour., 2007, 165, 509-516; G. L. Soloveichik, Annu. Rev. Chem. Biomolecul. Engine., 2011, 2, 503-527.
6. W. Hoopes, U.S. Patent 1,534,318. 1925-4-21.
7. N. Jayaprakash, S. K. Das, L. A. Archer. Chem. Commun., 2011, 47, 12610-12612.
8. D. Sadoway, G. Ceder, D. Bradwell. U.S. Patent 8,268,471[P]. 2008-2-21.
9. H. Kim, D. A. Boysen, D. J. Bradwell, et al. Electrochim. Acta, 2012, 60, 154-162; H. Kim, D. A. Boysen, T. Ouchi, et al. J. Power Sour., 2013, 241, 239-248; K. Wang, K. Jiang, B. Chung, et al. Nature, 2014, 514, 348-350.
10. M. Gaberšček, V. Napast, J. Moškon, et al. Acta Chim. Slov., 2015, 62, 796-804.
11. T. Hutter, W. A. C. Bauer, S. R. Elliott, et al. Adv. Func. Mater., 2012, 22, 2624-2631; K. Ma and J. Liu. Phys. Lett. A, 2007, 361, 252-256; R. Viskanta, D. M. Kim, C. Gau, Int. J. Heat Mass Transfer, 1986, 29, 475-485.
13